\documentclass[11pt]{article}
\usepackage{DGfest}
\usepackage{times}

\def\Journal#1#2#3#4{{#1} {\bf #2}, #3 (#4)}

\def\NPB{{\em Nucl. Phys.} B}
\def\PLB{{\em Phys. Lett.}  B}
\def\PRL{\em Phys. Rev. Lett.}
\def\PRD{{\em Phys. Rev.} D}

\def\NPBS{{\em Nucl. Phys. Suppl.} B}
\def\JETPL{\em JETP Letters}

\def\PPS{\em Progr. Phys. Suppl.}
\def\PPNP{\em Prog. Part.  Nucl. Phys.}
\def\JHEP{\em JHEP}
\def\ATMP{\em Adv. Theor. Math. Phys.}
\def\NPPS{\em Nucl. Phys. Proc. Suppl.}
\def\be{\begin{equation}}
\def\ee{\end{equation}}
\def\bea{\begin{eqnarray}}
\def\eea{\end{eqnarray}}

\begin{document}

\baselineskip 11.5pt
\title{MATTER OF RESOLUTION: FROM QUASICLASSICS TO FINE TUNING}

\author{ V.I. ZAKHAROV}

%\address{     MPI, F\"ohringer Ring 6, 80805, M\"unchen, Germany}
       % E-mail:xxz.mppmu.mpg.de

%\author{ C.D. AUTHOR2, E.F. AUTHOR3 }
%
\address{Dipartimento di Fisica, Universit\`a di Pisa, Largo Pontecorvo 3,\\
        I--56127 Pisa, Italy;\\
        MPI, F\"ohringer Ring 6, 80805, M\"unchen, Germany}

\maketitle\abstracts{
Recently, there appeared results of lattice measurements
in Yang-Mills theories
which indicate non-trivial dependences on the
lattice spacing of many observables.
In particular, volume occupied by
fermionic  zero modes  shrinks to zero in the continuum limit.
These results are  in apparent disagreement with  
quasiclassical models which assume that all the non-perturbative effects 
develop on the scale of $\Lambda_{QCD}$.
We emphasize that this kind of contradictions might be superficial
since the results in point depend in fact on the measurements procedure.
The lattice simulations correspond to 
measurements with high resolution
while the quasiclassical picture assumes poor resolution.
We will argue that the general trend is
that what looks quasiclassical in measurements with poor resolution
becomes fine tuned in measurements with fine resolution.
The main emphasis is on the topological fermionic modes and we argue that 
shrinking of the volume occupied by the modes could have been predicted theoretically.}

\section{Introduction}
%\subsection{Producing the Hard Copy}\ label{subsec:prod}

Some time ago it was demonstrated  \cite{adriano} 
that confining fields are soft, $A_{\mu}^a\sim \Lambda_{QCD}$. 
Namely, both confining and topological 
properties survive if one removes 
from the vacuum fields  large quantum fluctuations, $A_{\mu}^a\sim 1/a$,
where $a$ is the lattice spacing

More recently, however, there appeared measurements on original quantum field configurations
$\{A_{\mu}^a(x)\}$, without cooling, which demonstrate unexpected dependences on
the lattice spacing, or the ultraviolet cut off. 
In particular, the so called center vortices which appear to be responsible for confinement,
for references see \cite{review1} possess ultraviolet divergent non-Abelian action: \cite{gubarev}
\be\label{uv}
S_{vort}~\approx~0.54{(Area)_{vort}\over a^2}~~,
\ee
where $a$ is the lattice spacing. On the other hand, the area of the vortices is in physical units:
\be\label{area}
(Area)_{vort} ~\sim ~\Lambda^2_{QCD}V_{tot}~~,
\ee
where $V_{tot}$ is the volume of the lattice and references can be found in \cite{review1}.
Thus, according to measurements (\ref{uv}) and (\ref{area}) the confining fields 
are not soft but rather 
fine tuned, exhibiting dependences both on the UV and IR scales.

Most recently, it was also found  that the volume occupied by low-lying fermionic modes 
shrinks to zero
in the continuum limit. \cite{random} 
In more detail, the volume is defined in terms of the so called Inverse
Participation Ratio (IPR), see, e.g., \cite{gattringer}. One finds eigenfunctions of the Dirac equation, 
\be
D_{\mu}\gamma_{\mu}\psi_n~=~\lambda_n \psi_n~~,
\end{equation}
where the covariant derivatives $D_{\mu}$ are constructed on quantum vacuum configurations
of the gauge fields $\{A_{\mu}^a(x)\}$.
Furthermore, one introduces  $\rho_n(x)=(const)|\phi_n|^2$ normalized as
\be\label{powers}
\int d^4x \rho_n(x)~=~1~~.
\end{equation}
The IPR is defined then as
\be
(IPR)_n~\equiv~V_{tot}\int d^4x \rho_n^2(x)~\equiv~{V_{tot}\over (V_{mode})_{n}}~~,
\end{equation}
where $V_{tot}$ is the total volume and $(V_{mode})_{n}$
is the  volume occupied by the mode, by definition.
The observation is that 
for low-lying modes one has:
\be\label{shrinks}
\lim_{a\to 0}{V_{mode}}\sim ~(a\cdot \Lambda_{QCD})^{r}V_{tot},
\end{equation}
where $a$ is the lattice spacing and $r$ is a positive number of order unit.
In other words, the localization volume apparently shrinks to zero in the continuum limit $a\to 0$. \footnote{Independent evidence in favor
of shrinking volume of topologically nontrivial gauge-field configurations
was obtained in Ref. \cite{boyko}.}

All this accumulating evidence in favor of non-trivial dependences 
 on the lattice spacing  came unexpected because the common
belief is that all non-perturbative effects in Yang-Mills theories
are controlled by the scale of $\Lambda_{QCD}$. In particular,
the fermionic zero modes are a classical application of the instanton 
model, beginning with the seminal paper \cite{thooft}.

In an attempt to appreciate the result (\ref{shrinks}), let us first emphasize
that by measuring (IPR) and repeating or averaging   over such measurements 
we do not determine any matrix element.
Therefore, answers to such questions as
`what is the volume occupied by zero modes'  can well depend on
the measuring procedure itself. 
The latest results on the low-lying fermionic modes  \cite{random}
 correspond to measurements with high resolution.
Indeed, fixing a particular
 vacuum configuration $\{A_{\mu}^a(x)\}$ via Monte Carlo simulations  
 is equivalent to measuring 
all the fields with resolution $a$. 
On the other hand, the classical instanton calculus, see, e.g.,  \cite{thooft},
 applies to the case
when one specifies initial and final states and no measurements are performed in  between.
In other words, quasiclassical picture corresponds to measurements with poor resolution.

Using  examples, we will argue that
the general trend is that quasiclassical 
fields measured with high resolution
appear fine tuned.

Our main emphasis is on topological fermionic modes, section 3. 
We argue  that in fact the result (\ref{shrinks}) could have been predicted theoretically
and suggest bounds on the value of $r$:
\be\label{bounds}
1~\le~r~\le~3~~.
\ee
Section 2 is a kind
of introduction where we reiterate well-known facts about monopole clusters. The reason
is that similar structures arise in other cases which we will consider.
In Section 4 we emphasize the short-distance facet of the lattice strings mentioned
above, see Eqs (\ref{uv}) and (\ref{area}). In Section 5 we discuss briefly possible 
connection between strings and
fermionic modes.

\section{A case study: monopole clusters}

As a kind of introduction let us describe briefly a case where the
phenomenon which we call transition from quasiclassics to fine tuning
is well known and explicit. We mean description of the phase transition
to confinement in compact U(1) theory in terms of percolating cluster. \cite{polyakov}
In more general terms, we will recapitulate
a few points, in a non-systematic manner, 
from so called polymer approach to field theory, see, e.g., \cite{parisi}.
\footnote{Applications to phenomenology of lattice 
Yang-Mills theories have been discussed many times as well,
see \cite{vz}. }

\subsection{Percolating cluster}

In all the cases which we will consider fine tuned configurations 
are in fact percolating clusters of trajectories or surfaces.
Let us recall, therefore, the simplest model for percolating cluster.
Assume that probability of a given link to belong to 
a `trajectory'  is equal to $p$ and
is independent on other links. Then the probability to find a 
connected trajectory of length $L$ is
proportional to
\be\label{probability}
W(L)~\sim ~p^{L/a}N_L~\approx~exp{L\over a}\big(\ln p + c_{geom}\big)
\ee
where $N_L$ is the number of different trajectories of same length $L$
and $c_{geom}$ is the number of choices, on a given lattice to continue the
trajectory. For example, for hyper-cubic  lattice in 4d 
$$c_{geom}~=~\ln 7,$$ 
if we concentrate
on closed trajectories. 

If we start with very small $p$ and gradually increase it,
 we encounter a phase transition to percolation at
\be
p_{cr}~=~1/7~.
\ee
At this point, infinite-length trajectory is allowed and an infinite, or percolating cluster
emerges.
One of the central points is that even at $(p-p_{cr.})\equiv\epsilon\ge 0$ the percolating cluster is 
dilute as far as $\epsilon$ is small:
\be\label{dilute}
\theta_{link}\sim~(p-p_{cr})^{\alpha}~~,
\ee
where $\theta_{link}$ is the probability of a given link to belong to the infinite
cluster and the critical exponent is positive, $\alpha>0$.

\subsection{Compact U(1)}

The percolation picture can  readily be mapped onto a
free field theory (in Euclidean space), see, e.g. \cite{parisi}. 
As an example close to our subject consider theory with action 
\be
S~=~{1\over 4e^2}\int d^4x F_{\mu\nu}^2~~,
\ee
where $F_{\mu\nu}$ is the electromagnetic field. Then there exists a classical solution, Dirac monopole
\footnote{`Compactness' of U(1) is equivalent to postulating that the Dirac string
does not cost action.} with magnetic field ${\bf H}\sim (1/e){\bf r}/r^3$. The corresponding self energy
diverges in the ultraviolet:
\be \label{mm}
M_{mon}~=~{const\over e^2 a}~~,
\ee
where $a$ is the lattice spacing and the constant is known explicitly for a given regularization.
Then the action associated with the monopole trajectory is
\be\label{monact}
S_{mon}~=~{const \over e^2}{L\over  a}~~,
\ee
and the factor $exp(-S_{mon})$ can be replaced by the factor  $p^{L/a}$ in Eq (\ref{probability})
if we choose  $p=exp(-const/e^2)$, where the constant is the same as in Eq. (\ref{mm}).

Next steps are the same as in the preceding subsection. The phase transition to percolation of  monopoles
is simultaneously phase transition to confinement of external electric charges.

Note that the mapping of percolation theory into the free field theory can be completed by
evaluating path integral for
propagator of a particle with the classical 
action (\ref{monact}). The propagating mass, $m_{phys}$
turns to be:
\be\label{m2}
m^2_{phys}~\approx~{const^{'}\over a^2}\big({const\over e^2}-\ln 7\big)~~.
\ee
The central point is that only fine tuned theories with
\be
e^2-e^2_{crit}~\equiv~\epsilon~\sim~a^2m^2~
\ee
are interesting. Indeed, only in this case $m^2$ can be kept independent of the lattice spacing
with $a\to  0$.

\subsection{Fine tuning and short distances}

Keeping $m^2$ in Eq (\ref{m2}) finite corresponds to fine tuning between energy and entropy.
Both quantities are UV divergent while their difference is not. Indeed,
Eq (\ref{m2}) is an example of a general relation:\be\label{free}
(Free~energy)~=~(Energy)~-~(Entropy) ~~.
\ee
Only free energy is `physical' and it does not depend on the lattice spacing. One can wonder what is then use to follow fine
tuning between the energy and entropy if none of these quantities is of
direct physical meaning.

The answer to this question is that by observing fine tuning, as cancellation between two
UV divergent quantities, we check that the monopoles are indeed point-like. 
The polymer approach to field theory which we are in fact highlighting now 
turns relevant
to the monopole physics because monopoles are defined at short distances, in
terms of violations of Bianchi identities:
\be\label{topology}
j_{\mu}~\equiv \partial_{\nu}\tilde{F}_{\mu\nu}
~~,\ee
where  $j_{\nu}$ is the monopole current. Clearly enough, Bianchi identities 
can be violated only on singular fields.

Thus, in measurement with resolution of order lattice spacing $a$ percolating
monopole cluster is very dilute, see (\ref{dilute}) and there is fine tuning,
see (\ref{m2}). In the classical picture, on the other hand, the same
phenomenon of the monopole condensation is described as 
\be
<\phi_M>~\neq~0~~,
\ee
where $\phi_M$ is magnetically charged field. In the classical picture $<\phi_M>\neq 0$ and is the same
everywhere. This picture  corresponds to smearing of the dilute cluster seen at short distances.

\section{Topological fine tuning of fermionic modes}

\subsection{Tunneling time: QM case}

Rather recently, there was an extensive discussion, how long does it take for a particle
to traverse a (quantum-mechanical) barrier, for review see \cite{chiao}.
Below,
we will argue that the results obtained are a kind of analogy to what we are observing
now with localization of fermions in YM.

In more detail, the tunneling time is defined in the following way.
Consider the famous two-slit experiment. But put now a barrier on one of the two trajectories, 
so that classically light traveling along this trajectory cannot reach the
detector. Quantum mechanically, however, some light goes through and one does observe
the interference pattern on the screen. By analyzing the interference pattern, furthermore, 
one can deduce the time $\Delta t$ which the particle spent `under the barrier'.
The result is:
\be\label{speed}
\Delta ~t~=~ 0~~.
\end{equation}
No matter how strange the result might look at first sight, it was predicted
within QM. However, the interpretation is not of much interest for us here.

Note that the information on small probability of tunneling is not lost.
The point is that the intensity of light penetrating the barrier is small.
But, by observing the interference one picks up only the particles which
did go through the barrier. And it is on these, selected trajectories
that one observes instantaneous transition (\ref{speed}).

Two remarks are now in order. There is no contradiction, of course, between
the quasiclassical calculation of the probability of the barrier transition and
observation (\ref{speed}). To verify the quasiclassical calculation
one needs measurements with poor resolution, averaging
over long time intensity
of light penetrating the barrier. The instantaneous transition (\ref{speed})
is observed if the resolution is fine: by observing the interference we pick
up the particle  which went through the barrier.

The second point to mention, is that the result (\ref{speed}) allows for a nice,
although might be too naive interpretation. Namely, one can say that particle
under the barrier travels in `imaginary time'. And then it is only natural that by
measuring time, which is real, we cannot detect how long the particle
lived in the imaginary time.

\subsection{Lessons?}

In case of YM theory,  there are barrier transitions
described by instantons
and in the next section we will establish analogy between the QM example just described and 
measurements in the YM case.
Here we will emphasize that
the problem we are confronting now is not
calculation of quantum corrections to the probability of
the instanton transition.

Instantons determine the matrix element of transition between two topologically distinct vacua.
Keeping the exponential alone the probability of the tunneling is proportional:

\be\label{classical}
P_{tunneling}~\sim~\exp\big(-{8\pi^2\over g^2}\big) V_{tot}~,
\end{equation}
where $g^2$ is the (classical) coupling.
The quantum correction to this result on one loop level was 
fully calculated \cite{thooft}
and the result is 
\be\label{quantum}
P_{{tunneling}}~\sim~\exp\big(-{8\pi^2\over g^2(\rho^2)}\big) {d^{4}x d\rho\over \rho^{5}}~,
\end{equation}
where $\rho$ is the instanton size. 
Eq (\ref{quantum}) completes calculation of quantum corrections 
of the probability of the tunneling.
However, the standard instanton trajectory dominates (\ref{quantum})
provided that we observe only initial and final states which
are separated by a very long `time',
$$T\gg \rho~\sim~\Lambda_{QCD}^{-1}$$ There are no measurements in between.
Now, we will consider measurements 
with high resolution.
The matrix element (\ref{quantum}) does not change,
as far as the measuring procedure does not induce topologically non-trivial
transitions. Which we assume to be the case since 
measurements in YM case make simply manifest zero-point
fluctuations.

The problem which we will address now is how
to predict properties of the `instanton' trajectory  provided that the measurements are made 
on scale $a$ and
$$a\ll \Lambda_{QCD}^{-1}~~.$$ 
This is a problem different from
evaluating quantum corrections to (\ref{classical}).
Indeed, prediction (\ref{speed}) is not a result of quantum corrections to the
standard  quasiclassical calculation of the probability of the barrier transition
but an answer to a different question.

 \subsection{Protected matrix elements in the YM case}

Quantum-mechanical examples suggest  two tools
to translate the quasiclassical picture into the picture 
obtained with high resolution.
First, one can use matrix elements which are protected
against the ultraviolet noise. These should not depend on the resolution.
Second, we arrive at the principle
that tunneling time cannot be measured, even if the measurements are
made with high resolution. Combining the two observations we shall be 
able to derive fine tuning of the fermionic modes.

In case of the Yang-Mills theory, one measures 
two matrix element related to the low-lying fermionic modes. 
These are the topological susceptibility
(Witten relation),
\be\label{witten}
<Q^2_{top}>~\sim~\Lambda_{QCD}^4V_{tot}~~,
\end{equation}
and density of the near-zero modes (Banks-Casher relation):
\be\label{bc}
<\bar{q}q>_{quenched}~=~-\pi ~\rho(\lambda_n\to 0)~\sim \Lambda_{QCD}^3~.
\end{equation}
Both matrix elements (\ref{witten}) and (\ref{bc})
can be estimated within quasiclassical, instanton picture and measured
on soft-field configurations which obtained via cooling,
for review see, e.g., \cite{teper}.
Since perturbation theory does not contribute to (\ref{witten}), (\ref{bc})
one expects that these relations remains true in measurements with
high resolution. And, indeed, 
in both cases there is no dependence on the lattice spacing,
see  second paper in Ref. \cite{random}. \footnote{
There exist also `unprotected' matrix elements relevant
to the study of the topological properties of the vacuum.
Consider correlator of topological densities.
Perturbatively, the correlator is easy to calculate:
\be\label{pert}
<G\tilde{G}(x),G\tilde{G}(0)>~=~ -{144\over \pi^4 x^8}~~,
\end{equation}
and at short distances this is a valid approximation. 
From Eq (\ref{pert}) alone one concludes that local value of topological charge density
is of order
\be
G\tilde{G}(x)~\sim~{12\over \pi^2 (\Delta x)^4}~~,
\end{equation}
where $\Delta x$ is the resolution. Thus, distribution of the density of topological
charge is sensitive to the resolution.

Moreover, Eq (\ref{negative}) implies that that there can be no accumulation of
topological charge on a finite 4d volume. \cite{horvath}
Thus, structure of the density (of topological charge) distribution,
imposed by perturbation theory, looks as lumps of size of order $a^4$
of large charge density, $\sim 1/a^4$ with alternating sign. 
If one follows the chain of lumps of the topological charge of the same sign, 
the emerging structure seems to be 3d percolating volume.
\cite{horvath,ilgenfritz} It is worth emphasizing that this `3d' volume 
does not know anything about $\Lambda_{QCD}$ and has such a fractal 
dimension that `3d' volume fills in a finite part of the 4d volume.

To summarize, gross features of the topological charge density 
distribution \cite{horvath,ilgenfritz}
could well be explained by perturbation theory. For further conclusions it
would be very useful to have a `reference point' and perform similar measurements
of the density distribution in a trivial case of an Abilene gauge field.} 

\subsection{Tunneling volume: YM case}

Imagine that we decide to design an experiment in the YM case which is analogous 
to measuring tunneling time in QM. Then, naturally, we would proceed in
the following way.

Barrier transitions in YM theories are well known, that is, instantons. 
Thus, we would like to observe an instanton trajectory,
with good resolution and measure time (in 4d case, 4d volume in fact)
it takes.  
Well, it is not so simple since measurements on the trajectory introduce fields
of order
\be
(A_{\mu}^a)_{quant}~\sim ~1/a~~,
\end{equation}
while the quasiclassical fields in the physical vacuum are
of order
\be
(A_{\mu}^a)_{class}~\sim~\Lambda_{QCD}~\ll~a^{-1}~~.
\end{equation}

Thus, we have a huge noise. However, presumably, this noise is
topologically trivial. Then,  an ingenious way to suppress the noise 
is to concentrate on zero fermionic modes. Indeed,
the index theorem guarantees
\be\label{zeromodes}
n_{+}-n_{-}~=~\Delta Q_{top}~~.
\end{equation}
To derive this equation one integrates over  the whole volume.
However, in reality the integral is saturated by the regions
which dominate $|\psi_{0}|^2$.
Thus, to measure the volume occupied by the zero mode seems to be a
smart way to determine how long the instanton (barrier) transition takes 
place!

Moreover, there is a general relation \cite{sign}
\be\label{negative}
<G\tilde{G}(x),G\tilde{G}(0)>~<0~~,
\end{equation}
which follows from unitarity alone
and which is satisfied, of course, by the perturbative contribution. 
Note that for instantons the sign is opposite, 
\be \label{positiv}
<G\tilde{G}(x),G\tilde{G}(0)>_{instanton}~<0~~,
\ee
as far as the points $(0,x)$ 
are within one instanton. 
Thus, trying to separate instanton trajectory 
from the perturbative noise we are trying to pick up
a `non-unitary' contribution. 
 This is an analogy to the `imaginary vs real' time in case 
of the barrier transition in QM.

Since this point is crucial let us reiterate the argument in terms 
of dispersion relations. Introduce to this end the function $f(Q^2)$:
\be
\int d^4x \exp(iqx)\langle 0|T\{G\tilde{G}(x),G\tilde{G}(0)\}|0\rangle
\equiv f(Q^2)~~,
\ee
where $Q^2=-q^2>0$.
Then
\be
f(Q^2=0)~\sim~<Q^2_{top}>~>~0~~.
\ee
In the language of the dispersion relations, the positive sign
of  $f(Q^2=0)$ can be ensured only  by introducing a proper subtraction constant
in dispersion relations \cite{sign1}. Indeed, 
\be\label{negativ}
Im f(Q^2<0)~<~0~~,
\ee
and the dispersive contribution to $f(Q^2=0)$ would be  negative.

Thus, by measuring the volume occupied by zero modes we 
perform measurements on the `size' of a subtraction constant.
Naturally we reveal its local nature, manifested
by the vanishing of the volume occupied by zero modes
(which echo the shape of the nontrivial topological gauge field configuration).

Note that the low-lying modes appear fine tuned (see second paper in Ref. \cite{random})
in the sense that their volume
tends to zero while the energy stays stable. 
With the realization that non-trivial topological transitions correspond to
a subtraction constant we get an explanation of this phenomenon of the topological
fine tuning.
\footnote{One could expect that there exists also a more general mechanism for IPR
growing with $a\to~0$.
It is interesting to note that removing the topological modes seemingly removes all
the non-trivial values of IPR. \cite{gattnar,random} Thus there is no other
mechanism for fine tuning, but the topological one. }

\subsection{Dimensionality of the ''subtraction volume'' }

An open question is, to which manifold shrink the topological  modes.
Trying to get an answer, notice first that in the quenched approximation the fermionic modes are delocalized, i.e.,
$$V_{mode}\sim~V_{{tot}}~.$$
Indeed, if there were localized modes, then there existed mobility edge,
$\lambda_{mob}$ separating localized and delocalized modes.
Furthermore, there exists an obvious inequality:
\be\label{inequality}
~\lambda_{mob}^2~\le m^{2}.
\end{equation}
Indeed, by mass we understand the lowest eigenvalue which
corresponds to the ordinary plane wave, propagating through
the whole volume. The eigenfunction corresponding to the mass
is not localized. Localized states correspond to a kind 
of a bound state since the particle moves within a  finite volume.  

In the quenched approximation chiral fermions cannot acquire mass
by symmetry considerations, $m^{2}=0$
and Eq (\ref{inequality}) immediately implies that fermionic modes
 cannot be localized. 
 
 This means in turn that the modes extend through the whole volume
 and occupy at least a $d=1$ manifold:
 $$V_{{mode}}\sim~V_{tot}(a\cdot\Lambda_{QCD})^{r}~,~~r\le ~3~.$$
 In this way we come to the upper bound in Eq (\ref{bounds}).
 The lower bound on the value of $r$, see Eq (\ref{bounds}) follows from the observation
that the anti-unitary sign in (\ref{positiv}) corresponds to one of the coordinates, `time'
having extra factor of $\sqrt{-1}$. The extension of the instanton
in the direction of this coordinate cannot be measured.

Consider now the unquenched case. A non-vanishing mobility edge 
is still not allowed. Indeed, let us start with the Banks-Casher
relation,  see Eq. (\ref{bc})
which is true in the unquenched case as well. 
We will also assume that the density $\rho(0)$ is not vanishing.
Then it remains true that the near-zero modes constrained by
this relation are delocalized.
Indeed, recall  that near-zero modes
become zero modes in the limit of the infinite volume,
according to this relation. If these
'near-zero' modes were localized in a finite volume, they would
not be sensitive to the total volume which is much larger than the
localization volume. The rest of argumentation leading to
(\ref{bounds}) remains unchanged compared to the quenched case.

\subsection{Summary on the fermionic modes}

Let us summarize briefly the issue of the topological fermionic modes.

Low-lying modes fill in the window in the perturbative spectrum:
\be
0~\le~\lambda_n~\le~{\pi\over L}~~,
\ee
where $L$ is the lattice size. The low values of the eigenvalues is a reflection
of the topological nature of the modes. Their shape follows the shape of topologically
non-trivial gluonic fields.

Furthermore, we argued that topological transitions are in fact `anti-unitary'.
Therefore measuring the volume occupied by the topological  
 fermionic modes is like performing measurements on a subtraction constant
which is local.  That is why performing measurements
with better and better resolution we observe that the topological modes shrink to a 
vanishing
4d volume. This is a mechanism of topological fine-tuning: the volume of a state
tends to zero while its `energy' remains small.

Thus, with vanishing lattice spacing. $a\to 0$ the topological modes become a kind of
percolating cluster. Predicting of the fractal dimension of this cluster 
remains a challenge.
At this moment, we can only suggest bounds (\ref{bounds}).

\section{Lattice strings}

On the theoretical side, there is no direct relation   between the topological fermionic modes,
or topologically non-trivial gluonic fields  and confinement. 
On the other hand, the lattice data suggest strongly that
deconfinement phase transition and restoration of chiral symmetry
happen at the same temperature and
one is invited to speculate that both phenomena originate from the
same vacuum field configurations.

As for the confining fields, there are good reasons to believe that
they are represented by the center vortices, for review see \cite{review1}. 
In measurements with fine resolution vortices appear fine tuned,
see Eqs (\ref{uv}), (\ref{area}) and then we reserve 
the word `strings' for the vortices 
since they do look like thin 2d surfaces.  
As far as one thinks about chiral symmetry breaking as triggered by instantons,
confining fields look unrelated to the breaking of chiral symmetry.
However, now that we are beginning to understand that in measurements with
fine resolution fermionic modes are also becoming a kind of a cluster,
a close  connection  between strings and topologically non-trivial fields is no longer
ruled out.

We will outline the latest data 
\cite{morozov} on this connection in the next section.
In this section we will emphasize connection between 
lattice strings and strings discussed within continuum theory,
see, e.g.,
\cite{maldacena}.  This connection is revealed through measurements with
high resolution.

%Here we emphasize possible
%connection with continuum theories.
%Namely topological definition of the lattice strings matches
%definition suggested by continuum theories \cite{maldacena,ooguri}.
%in the next section we will briefly discuss conncetion between
%fermionic modes and strings.

\subsection{'t Hooft loop}

There are two external probes of the vacuum state in non-Abelian case,
that is heavy quarks and heavy monopoles. The heavy-quark potential
is related  to the vacuum expectation value of the Wilson line:
\be
\langle ~W~\rangle~\sim~\exp(-V_{Q\bar{Q}}(R)T)~,
\ee
while the heavy-monopole potential is related to  the 't Hooft loop:
\be
\langle ~H~\rangle~\sim~\exp(-V_{M\bar{M}}(R)T)~.
\ee
Note that heavy monopoles are defined not in terms of an Abelian subgroup but 
rather in terms of the center group.
We will concentrate on the $S(2)$ case and then the center 
group is $Z_{2}$. 

In the confining theory: \cite{thooft1}
\be
\lim_{R\to \infty}{V_{Q\bar{Q}}}~=~\sigma R~;~
\lim_{R\to\infty}{V_{M\bar{M}}}~=~const~~.
\ee
Generically, confinement of color is ensured by condensation of the
`magnetic degrees of freedom'. The same condensation explains also 
screening of the magnetic charges and 
flattening of $V_{M\bar{M}}$ at large distances.

What are magnetic degrees of freedom is a dynamical question.
Assume that the Wilson line is determined in terms of strings living in 5d space with non-trivial geometry which can be open on the Wilson line, see, e.g.,
\cite{maldacena}. Then by magnetic degrees of freedom 
it is natural to understand strings which can be open on the 't Hooft loop.
Such objects are indeed commonly introduced in the continuum theory,
see, e.g., \cite{ooguri}.

\subsection{$Z_{2}$ gauge theory}
  
 To make the guess on maagnetic, or dual strings more transparent consider analogy to $Z_{2}$
 gauge theory. In this theory, see, e.g. \cite{rebbi}, 
 one integrates over link variables
 $Z_{\mu}(x)=\pm 1$.  The corresponding plaquettes take
 values $P=\pm 1$ as well. The action is defined as 
 proportional to the total area of the negative plaquettes:
 \be
 S_{Z_{2}}~=~\beta A_{neg} ~.
 \ee
 For the Wilson loop in this theory one can derive:
 \be\label{wilson}
 \langle~W~\rangle_{Z_{2}}~\sim~\Sigma_{A_{C}} \exp(-\beta^{{*}}A_{C})
 ~~, \ee
 where the sum is taken over all the surfaces span on the Wilson contour ${C}$
 and $A_{C}$ is the area of a surface while
 the constant $\beta^{*}$ is defined in terms of the original constant $\beta$~~
 in the following way:
 \be\label{hooft}
 \beta^{*}~=~-\ln tanh{\beta\over 2}~~.
 \ee
 Similar representation exists for the 't Hooft line:
 \be
 \langle~H~\rangle_{Z_{2}}~\sim~\Sigma_{A_{H}}\exp(-\beta A_{H})~~.
 \ee
 Naively, the representations (\ref{wilson}) and (\ref{hooft}) 
 imply the area law both for the Wilson and 't Hooft lines. 
 And this would be in contradiction with the confinement criteria mentioned above. 
 
 The resolution of the paradox is that one of the representations (\ref{wilson}), 
 (\ref{hooft}) is actually formal for any value of the constant $\beta$.
 `Formal' means that the sum over the surfaces is in fact divergent since the 
 entropy factor for the surfaces overweighs the suppression due to the action.
 Consider, for example, $\beta$ small. Then $\beta^{*}$ is large and the sum
 (\ref{wilson}) can be approximated by the surface with smallest area.
 And this implies the area law
 for the Wilson line. The sum (\ref{hooft}) for the 't Hooft line  is, however, divergent. If the constant $\beta$ and its dual are equal to each other,
 $$\beta~=~\beta^{*}~,$$ 
 then there is a phase transition.
 
 Physics-wise,  this divergence implies that the corresponding surfaces
 can form an infinite percolating cluster.
  Generically, the logic here is the same  as  was used
 when we considered percolating monopole clusters.
 The surfaces which percolate in the vacuum 
 are closed surfaces which can be open on the 't Hooft line.
 
In the non-Abelian case, there are arguments that string representation is
valid for the Wilson and 't Hooft lines in terms of strings living 
in higher-dimensional spaces with non-trivial geometry,
for review and references see \cite{klebanov}.
Following the example of the $Z_{2}$ gauge theory, one can expect that
the strings which can be open on the 't Hooft loop are actually condensed
and there is percolating cluster of such strings in the vacuum state of Yang-Mills
theories. \footnote{To our knowledge this point, or specualtion has not been
made in the continuum-theory literature on the magnetic (dual) strings.}

\subsection{Finding strings}

The problem of finding
such strings in the vacuum seems to be a formidable task, even if they
exist.
It is amusing, therefore, that phenomenologically this problem has been 
(approximately) solved, for references see \cite{review1} rather
long time ago and without reference to the theoretical developments
mentioned aove.  These strings are center vortices. 

Algorithmically, the center vortices are defined in terms of a
$Z_{2}$ projection. Namely,
one replaces the original field configurations by the closest configuration
of the $Z_{2}$-fields,
\be
\{A_{\mu}^a\}~\to~\{Z_{\mu}(x)\}
\end{equation}
 and the closeness
of the two sets of the fields is defined in terms of the norm of the fields,
summed up through the whole lattice. The center vortices are then defined in terms
of negative plaquettes, evaluated in the $Z_{2}$ projection. \footnote{Namely, the center vortices
are defined on the dual lattice as unification of all the plaquettes orthogonal 
to the negative projected plaquettes on the original lattice, for details see \cite{review1}.}

There is ample evidence that central vortices defined in this
way are relevant, or responsible for the confinement. \cite{review1}
However, there are many other ways to introduce $Z_{2}$ projections.
Phenomenologically, only those projections are successful which find
a string of negative plaquettes introduced by hand on the lattice. \cite{faber} 
We are now in position to  argue that the theoretical meaning of this phenomenological observation is 
that such strings can be open on the 't Hooft loop.
\footnote{The 't Hooft loop is the trajectory of end points of
the Dirac string. The Dirac string, in turn pierces nagative plaquettes.}
 To our mind, it is quite a remarkable success of the theory. 

From our perspective, it is crucial that the lattice strings are fine tuned
\cite{vz1}.
Indeed, observations (\ref{uv}) and (\ref{area}) imply that
\be\label{quasi}
(String~tension)~=~(Bare~tension)~-~(Entropy)~\sim~\Lambda_{QCD}^{2}~,
\ee
while the bare tension and entropy are ultraviolet divergent. This is another
example of the relation (\ref{free}) which we discussed in connection with
the monopole clusters. Eq (\ref{quasi}) suggests existence of a quasiclassical
description of the strings as well.

\subsection{Summary on the lattice strings}

Lattice strings are fine tuned.  It is well known, see, e.g., \cite{ambjorn}
that fine tuning is the only way to realize strings (in Euclidean space)
quantum mechanically.
However, there is no consistent string theory in 4d. The 
quasiclassical image of the
same strings probably is produced by theories with extra dimensions
and non-trivial geometry.
In these notes, we gave only one example of possible relation between the lattice
strings and strings of the continuum theory. Namely, originally lattice strings
are defined in terms of projected fields and the definition is
far from being transparent. Now, one can appreciate theoretially,
which projections are indeed physical.

\section{Bringing parts 3 and 4 together}

In two preceding sections we considered topologically non-trivial
gluon fields and confining field configurations. Both, when measured with fine resolution appear to be percolating clusters,
or submanifolds of the 4d space. 

It is tempting to speculate that the two clusters finally merge with each other.
One can check this hypothesis through direct
measurements. \cite{morozov}
Namely one measures correlation
between intensity of topological fermionic modeds and lattice strings.
The correlation turns to be positive and strong.

In more detail, the vortices live on the dual lattice and one considers
the correlator of the points on the dual lattice, $P_{i}$, which belong to 
the strings with the value of $\rho(x)$ averaged over the vertices of
the 4d hypercube , $H$, dual to $P_{i}$, where $\rho_{\lambda}(x)$ is the density
of the topological modes introduced in Eq (\ref{powers}).
Thus, the correlator studied is defined as:
\be
 C_{\lambda}~=~{\Sigma_{P_{i}}\Sigma_{x \in H}\big(V\rho_{\lambda}(x)~-~
 <V\rho_{\lambda}(x)>\big)\over \Sigma_{P_{i}}\Sigma_{x\in H} ~I}
\ee
It was found that the correlator is positive and large numerically
for the topological modes. Also, the larger eigenvalue $\lambda$ is, the smaller is the correlator.

Moreover, the correlator depends on the number of the vortex
plaquettes, $N_{vort}$ attached to the point $P_{i}$. For low-lying modes,
roughly speaking one gets: \cite{morozov}
\be
C_{\lambda\sim ~0}~\approx~0.1 ~N_{vort}~~,
\ee
where $3\le N_{vort}\le 10$.

These results are in agreement with the hypothesis that the vortices are related 
to the chiral symmetry breaking.
 Further efforts are needed, however, to 
clarify the connection between 
the lattice strings and the chiral symmetry breaking. 
\footnote{Indirectly it has been known for some time that
vortices and zero modes are related to each other. Namely,
by removing vortices one eliminates in fact topological fermionic modes.
\cite{deforcrand,gattnar} However, by this procedure one actually
affects 3d volumes, not only strings. \cite{syritsyn} Now, there are direct measurements available on the correlation of the modes with vortices and 3d volumes. The correlation
turns positive in both cases.} 

\section{Conclusions}

We argued that quasiclassical and fine-tuned description can be dual to each other
and apply to measurements with poor and fine resolutions, respectively.
While the quasiclassical picture is more traditional and, therefore, intuitive
the fine-tuned picture directly addresses physics of short distances. 
A remote analogy is that finding fine-tuned objects with ultraviolet
divergent action relevant to non-perturbative physics is similar to 
finding point-like quarks
of the perturbative physics.

\section*{Acknowledgments}
These notes are written for a volume dedicated to Adriano DiGiacomo
on the occasion of his 70th birthday. It is an honor for me to contribute
to this volume, in view of the great impact of the works of A. Di Giacomo 
on the whole field of lattice gauge theories.

An analogy between  measurements of the tunneling time in QM and YM cases
was worked out together with A. Vainshtein. I would also like to acknowledge 
detailed discussion of the whole text with A.V. Kovalenko and M.I. Polikarpov.
I am thankful to A. Di Giacomo, Ch. Gattringer, J. Greensite, F.V. Gubarev, S. Khlebnikov, M.E. Shaposhnikov, L. Stodolsky
for useful discussions. 

The main points of the talk were formulated during my short visits to the
 Princeton University and University of Minnesota. I am thankful to
I.R. Klebanov, A.M. Polyakov and M.A. Shifman, A.I. Vainshtein for the hospitality
and discussions.


\section*{References}
\begin{thebibliography}{99}
\bibitem{adriano}
M. Campostrini, A. Di Giacomo, M. Maggiore, H. Panagopoulos, E. Vicari,
\Journal{\PLB}{225}{403}{1989};\\
M. Campostrini, A. Di Giacomo, H. Panagopoulos, E. Vicari,
\Journal{\NPB}{329}{683}{1990};\\
 A. Di Giacomo, M. Maggiore, S. Olejnik, \Journal{\NPB}{347}{441}{1990}.


\bibitem{review1}
J.~Greensite, \Journal{\PPNP}{51}{1}{2003}, [arXiv:hep-lat/0301023].

\bibitem{gubarev}
F.V. Gubarev {\it et al.}, \Journal{\PLB}{574}{136}{2003}, [arXiv:hep-lat/0212003];\\
 V.G. Bornyakov {\it et al.} , \Journal{\PLB}{537}{291}{2002}, [arXiv:hep-lat/0103032].


\bibitem{random}
C. Aubin {\it et al.} {\it ``The Scaling Dimension of 
Low Lying Dirac Eigenmodes And Of The Topological Charge Density''},
[arXiv:hep-lat/0410024];\\
F.V. Gubarev, S.M. Morozov, M.I. Polikarpov, V.I. Zakharov,
\Journal{\JETPL}{82}{343}{2005}, [arXiv: hep-lat/0505016];\\
Y. Koma {\it et al.},{\it ``Localization properties of the topological charge density
and the low lying eigenmodes of overlap fermions''}, [arXiv:hep-lat/0509164];\\
C. Bernard {\it et al.}, {\it ``More evidence of localization in low-lying Dirac spectrum''},
[arXiv:hep-lat/0510025].

\bibitem{gattringer}
Ch. Gattringer {\it et al.}, 
\Journal{\NPB}{617}{101}{2001} [arXiv:hep-lat/0107016].


\bibitem{boyko}
P.Yu. Boyko, F.V. Gubarev,
{\it `` On the continuum limitg of topological charge density distribution ''},
[arXiv:hep-lat/0602001] ;\\
P.Yu. Boyko, F.V. Gubarev, S.M. Morozov,
\Journal{\PRD}{73}{014512}{2006}, [arXiv:hep-lat/0511050].

\bibitem{thooft}
G. 't Hooft, \Journal{\PRD}{14}{3432}{1978}.

\bibitem{polyakov}
A.M. Polyakov, \Journal{\PLB}{59}{82}{1975};\\
T. Banks, R. Myerson, J.B. Kogut, \Journal{\NPB}{129}{493}{1977};\\
H. Shiba, T. Suzuki, \Journal{\PLB}{333}{461}{1994}, [arXiv:hep-lat/9404015].

\bibitem{parisi}
G. Parisi, {\it ``Statistical Field Theory''}, Addison-Wesley, (1987) Chapter 16.

\bibitem{vz}
 V.I. Zakharov, {\it in}  ``Moscow 2003, I. Ya. Pomeranchuk and physics at the turn of the century'',
p 177, [arXiv:hep-ph/0306262]; {\it Phys. Usp.} {\bf 47}, 37 (2004); \\
M.N. Chernodub, V.I. Zakharov, \Journal{\NPB}{669}{233}{2003}, [arXiv:hep-th/0211267].

\bibitem{chiao}
R.Y. Chiao, {\it `` Tunneling times and superluminality: a tutorial''},
[arXiv:quant-ph/9811019].

\bibitem{teper}
M. Teper, \Journal{\NPBS}{83}{146}{2000},
[arXiv:hep-lat/9909124].

\bibitem{sign}
E. Seiler, I.O. Stamatescu, {\it ``Some remarks on the Witten-Veneziamo
formula for the $\eta^{'}$ mass''}, MPI-PAE/PTh 10/87.

\bibitem{horvath}
I. Horvath {\it et al.}, \Journal{\PRD}{67}{011501}{2003}, [arXiv:hep-lat/0203027].


\bibitem{ilgenfritz}
E.-M. Ilgenfritz {\it et al.}, {\it ``Probing the topological structure 
of QCD vacuum with overlap fermions''},
[arXiv:hep-lat/0512005].


\bibitem{sign1}
E. Seiler,  \Journal{\PLB}{525}{355}{2002}, [arXiv:hep-th/0111125];\\
 M. Aguado, E. Seiler, \Journal{\PRD}{72}{094502}{2005},
[arXiv:hep-lat/0503015].

\bibitem{gattnar}
J. Gattnar {\it et al.}, \Journal{\NPB}{716}{105}{2005} [arXiv:hep-lat/0412032];\\
S. Solbrig {\it et al.}, {\it ``Topologically non-trivial field configurations: interplay of vortices
and Dirac  eigenmodes''}, [arXiv:hep-lat/0509052].

\bibitem{thooft1}
G. 't Hooft, \Journal{\NPB}{53}{141}{1979}.
\bibitem{morozov}
A.V. Kovalenko, S.M. Morozov, M.I. Polikarpov, V.I. Zakharov,
{\it `` On topological properties of vacuum defects in lattice Yang-Mills theories''},
[arXiv:hep-lat/0512036]. 

\bibitem{maldacena}
A.M. Polyakov, \Journal{\NPPS}{68}{1}{1998}, [arXiv:hep-th/9711002];\\
J. Maldacena, \Journal{\ATMP}{2}{231}{1998}, [arXiv:hep-th/9711200].

\bibitem{ooguri}
 E. Witten,  {\it Adv. Theor. Math. Phys.} {\bf 2} 253 (1998), 
 [arXiv:hep-th/9802150];\\
N. Itzhaki, J. M. Maldacena, J. Sonnenschein, Sh. Yankielowicz,
\Journal{\PRD}{58}{046004}{1998}, [arXiv:hep-th/9802042];\\
D.J. Gross, H. Ooguri, \Journal{\PRD}{58}{106002}{1998}, [arXiv:hep-th/9805129].

\bibitem{rebbi}
M. Creutz, L. Jacobs, C. Rebbi, \Journal{\PRL}{42}{1390}{1979}.

\bibitem{klebanov}
I.R. Klebanov, {\it ``QCD and string theory''}, [arXiv:hep-ph/0509087].

\bibitem{faber}
M. Faber, J. Greensite, S.Olejnik, D. Yamada, \Journal{\JHEP}{9912}{012}{1999},
[arXiv:hep-lat/9910033].

\bibitem{vz1}
V.I. Zakharov, {\it ``Dual string from lattice Yang-Mills theory''},
{\it AIP Conf.Proc.} {\bf 756}, 182 (2005),  [arXiv:hep-ph/0501011].

\bibitem{ambjorn}
J. Ambjorn, {\it ``Quantization of geometry''}, [arXiv:hep-th/9411179].

\bibitem{deforcrand}
Ph. de Forcrand, M. D'Elia,  \Journal{\PRL}{82}{4582}{1999}, [arxiv:hep-lat/9901020].

\bibitem{syritsyn}
A.V. Kovalenko, M.I. Polikarpov, S.N. Syritsyn, V.I. Zakharov,
\Journal{\PLB}{613}{52}{2005}, [arXiv:hep-lat/0408014].

%\bibitem{bornyakov}
%.~G.~Bornyakov {\it et al.}, \Journal{\PLB}{537}{291}{2002}, [arXiv:hep-lat/%0103032].

%\bibitem{giedt}
%. Ambjorn, J. Giedt, J. Greensite, \Journal{\JHEP}{0002}{033}{2000}
%[arXiv:hep-lat/9907021].


\end{thebibliography}
\end{document}